\documentclass[aps,prx,reprint,groupedaddress]{revtex4-2}

\usepackage{amsmath}
\usepackage{amssymb}
\usepackage{float}
\usepackage{graphicx}
\usepackage{dcolumn}
\usepackage{bm}
\usepackage{hyperref}
\usepackage{orcidlink}
\usepackage[ruled,vlined]{algorithm2e}
\usepackage{booktabs}
\usepackage{multirow}

\begin{document}


\title{\textbf{
Decoder- and Noise-Aware Optimization Framework for Clifford-Deformed Quantum Error-Correcting Codes
}}


\author{Julius Rye Bønnelykke\orcidlink{0009-0006-7580-6713}}
\email{julius.rye.boennelykke@nbi.ku.dk}
\author{Benjamin Joecker\orcidlink{0000-0003-0263-5440}}
\author{Peter Krogstrup\orcidlink{0000-0002-1930-8553}}
\author{Vincent Philippe Michal\orcidlink{0000-0001-8402-5077}}
\email{vincent.michal@nbi.ku.dk}
\author{Mark Kamper Svendsen\orcidlink{0000-0001-9718-849X}}
\email{mark.svendsen@nbi.ku.dk}

\affiliation{\href{https://nqcp.ku.dk/}{NNF Quantum Computing Programme}, \href{https://nbi.ku.dk/}{Niels Bohr Institute}, \href{https://ror.org/035b05819}{University of Copenhagen}, 2100 Copenhagen, Denmark}

\date{\today}

\begin{abstract}
Practical fault-tolerant quantum computing requires qubit-efficient quantum error correction (QEC).
Clifford deformations (CDs) modify the stabilizers of a QEC code to lower its logical error rate at no additional resource cost.
However, the number of possible CDs grows exponentially with the number of qubits, making exhaustive search infeasible beyond small codes.
In this work, we further show that under circuit level noise the optimal CD depends strongly on both the noise model and decoder.
To make systematic CD optimization possible, we introduce Evolution-Based Best-arm Identification (EBBI), a shot-efficient search algorithm that treats the circuit, noise model and decoder as a black box.
We benchmark EBBI’s performance on the rotated surface code,  under two strongly $Z$-biased circuit-level noise models with the PyMatching decoder.
At distance 3, EBBI finds the optimal CD using up to $\sim 1100\times$ fewer shots than random search.
At distance 5, when run at the target error rate, the median EBBI run converges to the CDs we believe to be optimal, while random search never converges.
Under a $Z$-biased SI1000-like noise model, EBBI discovers a new CD, the B-XZZX code, which we generalized to larger distances.
It differs from XZZX on only a fraction $\mathcal{O}(d^{-1})$ of the qubits at code distance $d$, yet its relative advantage grows from $\approx 5\%$ at $d = 3$ to $\approx 12\%$ at $d = 11$.
We thus show that the B-XZZX pattern found by EBBI can be systematically scaled to larger code distances.
Our work represents a first step towards scalable CD optimization for general QEC codes.
\end{abstract}


\maketitle

\section{Introduction}

Quantum error correction (QEC) is widely regarded as the leading route toward fault-tolerant quantum computation~\cite{ShorPhysRevA.52.R2493,Shor1996,gottesman2026surviving}, and finding efficient QEC schemes is at the forefront of current research~\cite{webster_pinnacle_2026, cain_shors_2026, babbush_securing_2026}.
There exist QEC codes with a range of performances and demands on hardware.
Among them, the most well-known and widely used is the surface code~\cite{Bravyi1998quantum, dennis2002topological, fowler2012surface, krinner_realizing_2022, Acharya2023SuppressingQubit, Acharya2025QuantumThreshold}.
The surface code encodes a single logical qubit into a two-dimensional code patch with code distance $d$ consisting of $O(d^2)$ physical qubits, representing a considerable overhead in physical hardware.
More recently, efficient high-rate quantum low-density parity-check (qLDPC) codes~\cite{BreuckmannPRXQuantum.2.040101, bravyi2024high, liang2025generalized} have attracted attention for their ability to encode many logical qubits (possibly with constant encoding rate) within a code, although with the additional requirement of long-range connections, realized either in hardware or by algorithmic routing \cite{yoder2025tour, gu_nearest-neighbour_2026, nixon_vine_2026}.
The above-mentioned codes rely on implementing stabilizer circuits~\cite{gottesman1997stabilizer} from which errors on qubits may be detected and corrected.\\

Beyond the development of new QEC codes, the performance of a QEC implementation can be improved systematically at a fixed code layout by exploiting the structure of the noise and errors affecting the individual physical qubits~\cite{Aliferis-PhysRevA.78.052331}.
Modifications of the surface code stabilizer circuits were proposed as a strategy to exploit biased error channels and optimize code performance without increasing qubit overhead~\cite{tuckett2019tailoring,BonillaAtaides2021TheCode}.
This was later generalized to the concept of Clifford deformations (CDs) in the context of the surface code \cite{dua2024clifford}, and then applied to a family of high-rate qLDPC codes \cite{Roffe2023biastailoredquantum}.
Structure in the errors affecting qubits is relevant to a range of experimental platforms, including qubits realized by superconducting circuits \cite{lescanne_exponential_2020,DarmawanPRXQuantum.2.030345}, spin qubits in semiconductor quantum dots \cite{HetenyiPhysRevA.109.032433}, spins of defects in silicon \cite{Higginbottom_optical}, neutral atoms \cite{Sahay-PhysRevX.13.041013}, and trapped ions \cite{Seis-PhysRevA.107.052417}.

Previous approaches to CD optimization fall into four categories: analytical constructions for a specific code family~\cite{tuckett2019tailoring, BonillaAtaides2021TheCode}; local rules that set the deformation of each qubit from its dominant error axis~\cite{tiurev2023correcting}; exhaustive search on the distance-3 rotated surface code (referred to as the surface code for the rest of this paper) combined with a study of random CD families, from which a translation-invariant CD for larger codes is constructed~\cite{dua2024clifford}; and fast optimization of a decoder-independent surrogate for CD performance under spatially non-uniform noise~\cite{yun2026computationally}.
Specific CDs such as XZZX and XY have been studied under circuit-level noise~\cite{DarmawanPRXQuantum.2.030345, PhysRevX.13.031007, etxezarreta2026leveraging}, but searches for optimal CDs have so far been carried out under code-capacity noise with near-optimal decoding~\cite{dua2024clifford} or under phenomenological noise with an objective that ignores the decoder~\cite{yun2026computationally}, with only a limited circuit-level test in the latter.
Exhaustive search with the decoder in the loop accounts for the decoder by construction, but its cost grows exponentially with the number of qubits, while the scalable alternatives (local rules, random-family analysis and surrogates) are decoder-independent.
This lack of a method that is both scalable and decoder-aware has so far prevented a systematic assessment of the gains CDs can achieve under circuit-level noise with practical decoders.

Here, we introduce Evolution-Based Best-arm Identification (EBBI), a scalable optimization framework for identifying high-performing CD QEC codes under general circuit-level noise models and decoders.
The key observation underlying EBBI is that small changes to the CD patterns produce similar logical error rates $P_L$.
This allows information gathered from one configuration to guide the search toward promising regions of the exponentially large configuration space.
We evaluate EBBI using realistic circuit-level noise models suitable for a range of $Z$-biased qubits \cite{etxezarreta2026leveraging} and practical decoding based on PyMatching \cite{Higgott2025sparseblossom}.
Across a range of surface-code instances under circuit-level noise, EBBI consistently identifies CDs that either match or outperform previously proposed constructions under the considered noise models and decoding strategies while requiring several orders of magnitude fewer QEC evaluations than random search.
In particular, the search strategy allowed us to identify code boundary effects for our $Z$-biased SI1000-like noise-model [Sec.~\ref{sec:setup}] not reported so far [Sec.~\ref{sec:generalization}].
These boundary effects turn out to be relevant at large code distances, as scaling up the code while maintaining the structure found at low distance consistently outperforms previously identified codes at larger distances.
Our results establish a scalable CD optimization as a practical tool for tailoring quantum error-correcting codes to hardware-specific noise characteristics.\\

The structure of this paper is as follows.
In Section~\ref{sec:clifford_deformations} we recall the method of CD of a QEC codes.
In Section~\ref{sec:ebbi} we describe the EBBI algorithm as applied to the CD of QEC codes.
Then, we present in Section~\ref{sec:results} the optimal CD found with the EBBI framework, benchmark the algorithm's performance, and show how the method can be extended to large codes based on codes discovered at low distance.
We discuss and give an outlook in Section~\ref{sec:discussion}.

\begin{figure}[h!]
    \includegraphics[]{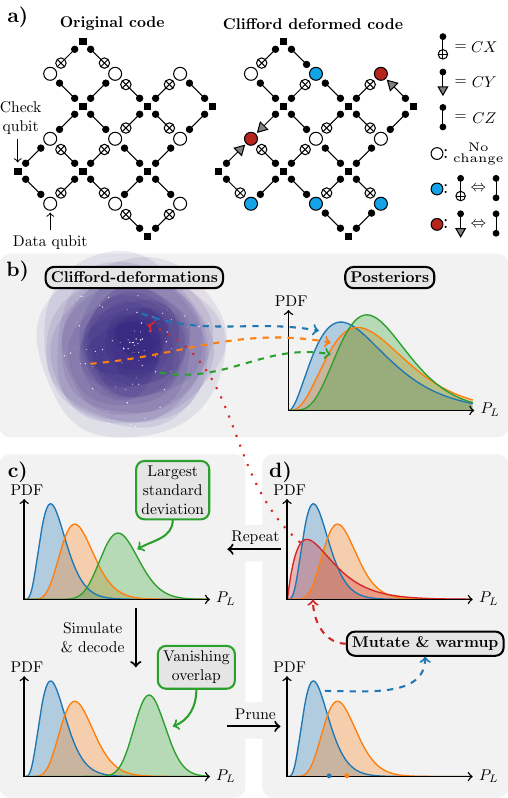}
    \caption{
        \textbf{a)} Example Clifford deformation (CD) on the distance-3 surface code.
        Data qubits are represented by filled circles and check qubits by black squares.
        Connections between data and check qubits represent the applied two-qubit gates.
        The color of a data qubit indicates the specific change made to its connected gates.
        \textbf{b,c,d)} Outline of the Evolution-Based Best-arm Identification (EBBI) method.
        \textbf{b)} \textit{Initialization:} Uniformly pick $K$ random CDs from $\mathcal{A}^n$ and simulate each until $W$ logical errors have been observed.
        The posterior over the logical error rate $P_L$ is modeled as a beta distribution.
        \textbf{c)} \textit{Sampling:} At each step, the posterior with the largest standard deviation is simulated and decoded $1.25S$ additional times with $S$ the number of shots already used for that candidate.
        \textbf{d)} \textit{Prune and re-populate:} Any posterior satisfying Eq.~\eqref{eq:prune_condition} is pruned and replaced by a mutation of the current best CD, selected by drawing one random number from each surviving posterior and taking the lowest; the mutation is warmed up similarly to panel b.
        If no CD satisfies Eq.~\eqref{eq:prune_condition}, EBBI returns to panel c.
    }
    \label{fig:ebbi}
\end{figure}

\section{Clifford deformations}\label{sec:clifford_deformations}

CDs are a technique for modifying the stabilizers of a QEC code without changing the code distance or the local connectivity map~\cite{dua2024clifford}.
The main advantage of CDs is that under biased noise, they can significantly reduce $P_L$ without any added hardware cost.
This is exemplified by the XZZX deformation on the surface code~\cite{BonillaAtaides2021TheCode}, which applies the Hadamard $H$ deformation on alternating qubits and, under biased noise, leads to an increased threshold of the code.
Ref.~\cite{dua2024clifford} showed that under symmetric $p_X=p_Y$ $Z$-biased code-capacity noise, the $P_L$ of the optimal distance-3 surface code CD was two orders of magnitude lower than that of the undeformed surface code.
Due to the exponential size of the search space, finding the optimal CD is a non-trivial problem and depends on both the noise model and the decoder.

From a code-capacity or phenomenological noise model perspective, CDs can be viewed in two ways:
\begin{enumerate}
    \item\textbf{The Heisenberg picture:} A deformation $U$ on a given data qubit of the code will change the part of the Pauli stabilizer $S$ that acts on that data qubit into $S' = USU^\dagger$.
    \item\textbf{The Schrödinger picture:} The stabilizers stay the same, but the noise on the qubit $(p_X,p_Y,p_Z)$ is permuted.
    The Hadamard $H$ deformation would give $(p_Z,p_Y,p_X)$ swapping the Pauli-X and Z probabilities.
\end{enumerate}
We note in passing that when circuit-level noise is considered, the Schrödinger-picture analog formally breaks down.
There exist 6 possible deformations per qubit, including no deformation $I/(p_X,p_Y,p_Z)$, $H/(p_Z,p_Y,p_X)$, $\sqrt{Z}/(p_Y,p_X,p_Z)$, $H\sqrt{Z}/(p_Y,p_Z,p_X)$, $\sqrt{Z}H/(p_Z,p_X,p_Y)$, and $H\sqrt{Z}H/(p_X,p_Z,p_Y)$ with $I$ the single-qubit identity gate, $H$ the Hadamard gate, and $\sqrt{Z}$ the $\frac{\pi}{2}$ $Z$-rotation gate.
Since these can be picked independently on each data qubit, a QEC code with $n$ data qubits will have $|\mathcal{A}_6^n|=6^n$ possible CDs with $\mathcal{A}_6=\{I,\,H,\,\sqrt{Z},\,H\sqrt{Z},\,\sqrt{Z}H,\,H\sqrt{Z}H\}$ the full CD set.
An example CD is shown in Fig.~\ref{fig:ebbi}a.

Symmetry in the noise model can reduce the number of unique CDs.
For example, in the code-capacity or phenomenological noise model regime, assuming $p_X=p_Y$ makes the 6 deformations pairwise identical: $I\!\sim\!\sqrt{Z}$, $H\!\sim\!\sqrt{Z}H$, and $H\sqrt{Z}H\!\sim\!H\sqrt{Z}$, resulting in a reduced CD set $\mathcal{A}_3=\{I,\,H,\,H\sqrt{Z}H\}$ with $|\mathcal{A}_3^n|=3^n$ CDs.
This becomes a bit more complicated in the circuit-level noise model regime, where single-qubit gates from circuit compilation [Appendix~\ref{app:circuit_construction}] will transform a $Z$-error into an $X$- or $Y$-error.
It can, however, be shown that the reduction from 6 to 3 deformations still holds in the case considered here [Appendix~\ref{app:cd_reduction}].
This reduces the search space from $6^9$ to $3^9$, matching the case Ref.~\cite{dua2024clifford} considered.
Symmetries thus have the ability to vastly reduce the total search space, although the space still remains exponentially large in the number of qubits $n$.

\section{Evolution-Based Best-arm Identification}\label{sec:ebbi}

EBBI aims to identify the optimal CD for an arbitrary Pauli-stabilized QEC code, biased noise model, and decoder using as few simulations as possible.
EBBI is a hybrid between a best-arm identification \cite{audibert2010best} algorithm and an evolutionary algorithm, as best-arm identification usually assumes a small, limited set of options, which is not the case here.
The advantage of EBBI is that it obtains all information about the circuit implementation, noise model, and decoder assumptions through simulation steps.
This allows it to work with any decoder, including those whose underlying assumptions are not known, as would be the case with machine learning decoders.
Below we outline the three steps EBBI goes through: \emph{initialization}, \emph{sampling}, and \emph{prune and re-populate}.\\

\paragraph*{Initialization} An active Clifford-deformation set $\mathcal{A}$ is selected.
This could be the full set $\mathcal{A}_6$, the reduced set $\mathcal{A}_3$, or some other subset of $\mathcal{A}_6$; the search space is then $\mathcal{A}^n$.
EBBI uniformly samples $K$ random CDs from $\mathcal{A}^n$ as candidates.
Each candidate CD undergoes a warm-up phase, in which simulations are run until $W$ logical faults have been observed.
The true $P_L$ of the CD is modeled as a beta distribution with parameters $\alpha=\mathrm{\#LogicalFaults}+1$ and $\beta=\mathrm{\#SuccessfulDecodings}+1$ [Fig.~\ref{fig:ebbi}b].

\paragraph*{Sampling} This step aims to reduce the uncertainty in the candidates $P_L$.
EBBI selects the candidate distribution with the largest standard deviation and samples it until its standard deviation has been reduced by a factor of $1.5$, requiring $S\cdot(1.5^2-1)=1.25S$ additional shots, where $S=\alpha+\beta-2$ is the number of shots already used on the candidate $\mathrm{CD}$ [Fig.~\ref{fig:ebbi}c].
The factor of $1.5$ is a hyperparameter of the model that can be fine-tuned: too small a value would trigger the simulation step too often and result in unnecessary overhead, while too large a value would risk wasting sampling steps on candidates that are later pruned out.

\paragraph*{Prune and re-populate} If the $\gamma$ lower bound of the $k$'th CD, $\mathrm{CD}_k$, after resampling exceeds the $1-\gamma$ upper bound of at least one other CD:
\begin{equation}
    F^{-1}_{\operatorname{Beta}(\alpha_k,\beta_k)}(\gamma)
    >
    \min_{j\neq k}
    F^{-1}_{\operatorname{Beta}(\alpha_j,\beta_j)}(1-\gamma),
    \label{eq:prune_condition}
\end{equation}
with $F^{-1}_{\operatorname{Beta}(\alpha,\beta)}$ the quantile function of a beta distribution with parameters $\alpha$ and $\beta$, $\mathrm{CD}_k$ is deemed unlikely to be optimal and is pruned; otherwise it is kept and this step is repeated.
We use $\gamma = 2.5\%$ (corresponding to non-overlap of the central $95\%$ credible regions) and observe good performance [Fig.~\ref{fig:Results}].
Too small a $\gamma$ is expected to waste valuable shots, while too large a $\gamma$ risks pruning out good candidates.

Once a CD is pruned via Eq.~\eqref{eq:prune_condition}, it is replaced by a new candidate.
Intuitively, two CDs that differ in deformation on only a single qubit, referred to as neighboring CDs, should have similar error rates, which motivates a mutation strategy.
This correlation in neighboring CDs has been validated in the distance-3 surface code and shown to have Pearson correlation of $\rho=0.83$ and $\rho=0.82$ for the two noise models that we are considering [Appendix~\ref{app:corr}], and from the above argument we would expect $\rho$ to increase with larger code distance as the relative change becomes smaller; EBBI's performance on the distance-5 surface code [Fig.~\ref{fig:Results}b,d] is consistent with this.
Based on this observation, the current best CD, CD\textsubscript{best}, is identified by drawing a single sample from each candidate's beta distribution and selecting the CD with the smallest sampled value, akin to Thompson sampling~\cite{william1933likelihood}.
The new candidate CD is then generated by, on each qubit, either picking a random deformation uniformly from $\mathcal{A}$ with probability $\mathcal{M}$ or copying the deformation on that qubit from CD\textsubscript{best} with probability $1-\mathcal{M}$, keeping the total number of active candidates fixed at $K$.
If the mutation has already been used, this step is repeated with $\mathcal{M}$ increased by $1\%$ until an unseen CD is found; if $\mathcal{M}$ exceeds $1$, no new CD will be added and EBBI continues with one fewer active candidate.
The new candidate undergoes the same warm-up process described above, before EBBI returns to the sampling step.

EBBI terminates once a shot budget or a wall-clock time limit is reached, at which point it returns the candidate with the lowest mean $\mu$ plus standard deviation $\sigma$.
We use $\min(\mu+\sigma)$ rather than $\min(\mu)$ to bias the returned CD towards candidates with good statistics (i.e., low remaining uncertainty in the $P_L$ estimate); this choice was observed to substantially reduce the false-positive rate.
See Algorithm~\ref{alg:ebbi} in the appendix or the full code and raw results on Zenodo \cite{EBBI_code}.

\section{Results}\label{sec:results}

\subsection{Setup}\label{sec:setup}

While EBBI will work for any stabilizer code, we here limit our focus to the surface code.
This is done for two reasons: 
(1) Several CDs of the surface code are already known, including the XZZX~\cite{BonillaAtaides2021TheCode} and XY~\cite{tuckett2019tailoring} CD, and it is the code that was studied in Ref.~\cite{dua2024clifford}, which introduced the generalization of CDs.
(2) The surface code has only a single logical qubit, meaning that performance improvement can be uniquely measured by the single logical qubit $P_L$. This is unlike codes with multiple logical qubits co-encoded in the same physical qubits where tradeoffs between the different encoded qubits $P_L$ might need to be considered.

Simulations are carried out in Stim~\cite{Gidney2021Stim:Simulator} with a circuit-level biased noise model.
The Stim circuits are compiled down to single-qubit rotations, CZ gates, and initialization and measurement in the Z-basis [Appendix~\ref{app:circuit_construction}].
The reported $P_L$ is determined from simulating and decoding the same circuit in both the X- and Z-basis under the same noise seed (circuit-level errors will be the exact same in both runs) and comparing the two results to determine which logical Pauli error occurred, giving us $P_L^I$ (no logical error), $P_L^X$, $P_L^Y$, $P_L^Z$ with $P_L = P_L^X + P_L^Y + P_L^Z$; we will refer to this as a single shot as the two simulation runs use the same seed and thus experience the same errors at the same locations.
The hybrid biased-depolarizing (HBD) noise model~\cite{etxezarreta2026leveraging} is used as a baseline.
All operations in the HBD noise model use the same error probability $p$.
However, on real devices, simple operations such as single-qubit gates will have lower error rates than more complicated operations like two-qubit gates.
To get closer to realistic noise while remaining platform agnostic, we have changed the relative noise terms to match the commonly used SI1000 noise model~\cite{gidney2021fault} and will refer to this modified HBD noise model as HBD-SI1000.
Both noise models are summarized in Table~\ref{tab:noise_models}.
Both noise models also satisfy the conditions that allow us to focus on the $\mathcal{A}_3$ deformation set instead of the full $\mathcal{A}_6$ set.

\begin{table}[tb]
    \centering
    \begin{ruledtabular}
    \begin{tabular}{lcc}
        Operation & HBD & HBD-SI1000 \\ 
        \hline 
        1Q gates & $\mathrm{DEP}_1(p)$ & $\mathrm{DEP}_1(p/10)$ \\ 
        $CZ$ & $\mathrm{Bias}_2(p,\eta)$ & $\mathrm{Bias}_2(p,\eta)$ \\ 
        $\mathrm{Init}_Z$ & $\mathrm{X\text{-}error}(p)$ & $\mathrm{X\text{-}error}(2p)$ \\ 
        $\mathrm{Meas}_Z$ & $\mathrm{X\text{-}error}(p)$ & $\mathrm{X\text{-}error}(5p)$ \\ 
        Data idle & $\mathrm{Bias}_1(p,\eta)$ & $\mathrm{Bias}_1(2p,\eta)$ \\ 
    \end{tabular}
    \end{ruledtabular}
    \caption{
        Implementation of hybrid biased-depolarizing (HBD)~\cite{etxezarreta2026leveraging} and the SI1000-like HBD (HBD-SI1000) as a function of the noise model strength $p=p_X+p_Y+p_Z$ and phase-bias $\eta=p_Z / (p_X+p_Y)$.
        See Appendix~\ref{app:noise_details} for details.
    } 
    \label{tab:noise_models}
\end{table}

The decoders PyMatching~\cite{Higgott2025sparseblossom}, BeliefMatching~\cite{PhysRevX.13.031007}, BP-OSD~\cite{roffe_decoding_2020}, Tesseract~\cite{beni2025tesseractdecoder} and the CudaQ tensor network (CudaQ-TN) decoder~\cite{cudaq, cudaqx} were tested on the CSS (undeformed code), XZZX, XY and B-XZZX [Sec.~\ref{sec:generalization}, Fig.~\ref{fig:generalization}a,b] CDs on both the distance-3 and distance-5 surface codes [Appendix~\ref{app:decoder_dependency}].
In all cases, Tesseract or CudaQ-TN gave the lowest $P_L$, followed by BP-OSD, BeliefMatching and finally PyMatching [Fig.~\ref{fig:decoder_ranking} in Appendix~\ref{app:decoder_dependency}].
PyMatching performed particularly badly on the XY surface code.
We attribute this to the $H\sqrt{Z}H$ deformation, which converts the dominant $Z$-errors into $Y$-errors that trigger detectors in both the $X$- and $Z$-type decoding graphs; PyMatching requires these errors to be decomposed into independent edges, discarding the correlation between them.
Consistent with this, BeliefMatching, which reintroduces such correlations~\cite{PhysRevX.13.031007}, improves $P_L$ more for the XY CD than for any other CD in all four settings [Fig.~\ref{fig:decoder_ranking} in Appendix~\ref{app:decoder_dependency}].
However, PyMatching was $113$--$413\times$ faster than the other decoders on a single CPU core, and $75$--$97\times$ faster than CudaQ-TN on an NVIDIA H200 GPU [Table~\ref{tab:decoder_runtime} in Appendix~\ref{app:decoder_dependency}].
At the $\sim 10^{11}$ shots per run used in Fig.~\ref{fig:Results}, this corresponds to $\sim 110$ core-hours per distance-5 run under HBD noise with PyMatching, compared with $\sim 22\,000$--$45\,000$ core-hours with BeliefMatching, BP-OSD, or Tesseract [Table~\ref{tab:decoder_runtime} in Appendix~\ref{app:decoder_dependency}].
We therefore use PyMatching to benchmark both EBBI and random search.

Estimating $P_L$ reliably requires $\mathcal{O}(1/P_L)$ shots, which becomes expensive as larger QEC codes are considered.
To mitigate this, we propose that during the search phase of either EBBI or random search we artificially increase the physical error rate from $p\rightarrow p_\mathrm{search}\geq p$.
As $p_\mathrm{search}$ gets closer to the code threshold, the code distance will have a diminishing effect on $P_L$, making it easier to reliably identify good CDs with a limited number of shots.
Based on an exhaustive evaluation of all $3^9$ distance-3 surface code $\mathcal{A}_3$-CDs at $p\in\{0.1\%,0.25\%,0.5\%\}$ and $\eta=500$, the ordering does not change significantly, as indicated by a Pearson correlation between $\rho=0.9975$ and $\rho=0.9998$ over all CDs, and between $\rho=0.9355$ and $\rho=0.9970$ over the top 100 CDs; HBD was generally observed to have higher correlation than HBD-SI1000 [Fig.~\ref{fig:Corr_HBD_MWPM} and Fig.~\ref{fig:Corr_MHBD_MWPM} in Appendix~\ref{app:p_search_corr}].

For testing, EBBI is compared against random search~\cite{dua2024clifford}, which uniformly and without replacement picks CDs from $\mathcal{A}^n$.
We run EBBI with $K=10$ candidates at a time, and sample each new candidate until $W=500$ logical faults have been observed.
We use a pruning threshold of $\gamma=2.5\%$ and a mutation rate of $\mathcal{M}=\frac{43\%}{n}$, which results in a single-qubit mutation $80\%$ of the time and a multi-qubit mutation $20\%$ of the time.
If the mutated CD matches one already used, it is discarded and resampled with $\mathcal{M}$ increased by $1\%$.
As argued in Appendix~\ref{app:cd_reduction}, we can limit our search to $\mathcal{A}_3$ instead of the full $\mathcal{A}_6$ set, reducing the number of possible CDs from $6^n\rightarrow3^n$.
Random search runs using a fixed number of shots per CD (listed in Table~\ref{tab:rs_settings}), which has been chosen to balance a reliable estimate of the $P_L$ of each CD against exploring a large number of CDs.

\begin{table}[tb]
    \centering
    \begin{ruledtabular}
    \begin{tabular}{lrr}
        $p_\mathrm{search}$ & Distance-3 surface code & Distance-5 surface code \\
        \hline
        0.001     & 1\,000\,000    & 10\,000\,000        \\
        0.0025    & 400\,000       & 4\,000\,000         \\
        0.005     & 200\,000       & 2\,000\,000       \\
    \end{tabular}
    \end{ruledtabular}
    \caption{
        Number of shots used by random search per CD, as a function of noise model strength during the search phase $p_\mathrm{search}$ and code distance of the surface code.
    }
    \label{tab:rs_settings}
\end{table}

\subsection{Performance}\label{sec:performance}

\begin{table}
    \centering
    \begin{ruledtabular}
    \begin{tabular}{lcc}
        & HBD & HBD-SI1000 \\ 
        \hline
         CSS ($d=3$) & $3.4\times10^{-3}$ & $4.4\times10^{-3}$ \\
         CSS ($d=5$) & $6.5\times10^{-4}$ & $7.4\times10^{-4}$ \\
         XZZX ($d=3$) & $\mathbf{2.6\times10^{-3}}$ & $3.9\times10^{-3}$ \\
         XZZX ($d=5$) & $\mathbf{3.5\times10^{-4}}$ & $5.4\times10^{-4}$ \\
         XY ($d=3$) & $5.4\times10^{-3}$ & $6.5\times10^{-3}$ \\
         XY ($d=5$) & $1.3\times10^{-3}$ & $1.2\times10^{-3}$ \\
         B-XZZX ($d=3$) & $3.3\times10^{-3}$ & $\mathbf{3.7\times10^{-3}}$ \\
         B-XZZX ($d=5$) & $4.6\times10^{-4}$ & $\mathbf{5.1\times10^{-4}}$ \\ 
    \end{tabular}
    \end{ruledtabular}
    \vspace{1ex}
    \small
    \caption{
        Comparison of surface code $P_L$ for different CDs and noise models considered in this work.
        For each noise model and code distance the score of the best performing code is highlighted in the table.
        The $P_L$ for the other mentioned decoders along with their uncertainties are reported in Table~\ref{tab:decoder_dependence} in Appendix~\ref{app:decoder_dependency}.
    }
    \label{tab:PyMatching_LERs}
\end{table}

\begin{figure*}[t]
    \includegraphics[]{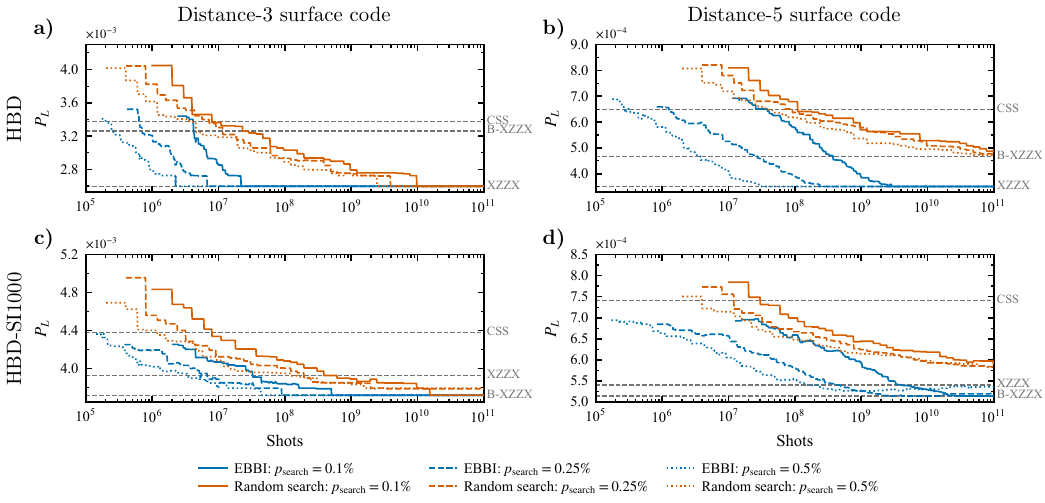}
    \caption{
        Median performance over 25 independent runs of EBBI and random search as a function of shot count.
        The $P_L$ of specific CDs of interest [Table~\ref{tab:PyMatching_LERs}] have been plotted as horizontal lines for comparison.
        The spread over all runs and over the central 15 out of 25 runs can be found in Fig.~\ref{fig:results_ci} in Appendix~\ref{app:results_ci}.
        \textbf{a-b)} use the HBD noise model~\cite{etxezarreta2026leveraging}.
        \textbf{a)} uses the distance-3 surface code with 3 syndrome extraction rounds.
        During the runtime of EBBI and random search, the noise model used noise strength $p_\text{search}$ and $\eta=500$, while the plotted points use $p=0.1\%$ and $\eta=500$ with an estimated $P_L$ from $10^8$ shots.
        Reference lines for the $P_L$ of the CSS (undeformed code), XZZX~\cite{BonillaAtaides2021TheCode} and B-XZZX (Fig.~\ref{fig:generalization}a,b) deformations are plotted, with the $P_L$ of the XY~\cite{tuckett2019tailoring} deformation lying outside the upper limit of the y-axis.
        \textbf{b)} same as panel a but for the distance-5 surface code with 5 syndrome extraction rounds.
        \textbf{c-d)} same as panels a, b but under our HBD-SI1000 noise model, where each operation uses a different noise strength, see Table~\ref{tab:noise_models}.
    }
    \label{fig:Results} 
\end{figure*}

Across all cases tested (each case is the median $P_L$ over 25 independent runs), EBBI shows a significant speedup over random search.
For the distance-3 surface code, EBBI finds the optimal CD in almost all cases [Table~\ref{tab:method_results} in Appendix~\ref{app:results_ci}] (confirmed by an exhaustive search over all $3^9$ $\mathcal{A}_3$-CDs [Appendix~\ref{app:exhaustive}]).
    Under $p_\mathrm{search}=p=0.1\%$ and HBD noise, the median EBBI run reaches the optimum using $\sim450\times$ fewer shots than random search [Fig.~\ref{fig:Results}a], and $\sim30\times$ fewer under HBD-SI1000 [Fig.~\ref{fig:Results}c].
    Increasing $p_\mathrm{search}$ to $0.5\%$ expands this gap to $\sim1100\times$ under HBD noise.
    Under HBD-SI1000 noise with $p_\mathrm{search}\in\{0.25\%,0.5\%\}$, the random search median never reaches the optimal $P_L$ for two reasons:
    (i) each random search run samples CDs without replacement and stops once it repeatedly fails to draw a CD it has not previously used, so in practice each run covers only $\approx78\%$ of the $3^9$ CDs; and
    (ii) its per-CD $P_L$ estimates are too noisy to resolve the separation between the optimal and second-best CDs, which is about one standard deviation of the random search estimate.
For the distance-5 surface code, no random search run converged: finding the optimum by uniform sampling requires on the order of $3^{25}\approx10^{12}$ CDs, each using $10^7$ shots, which is far beyond our simulation budget.
    Under HBD noise, the EBBI median reaches the XZZX deformation for all three $p_\mathrm{search}$ settings [Fig.~\ref{fig:Results}b]; we believe it to be optimal, as no CD explored by EBBI or random search beats it.
    Under HBD-SI1000 noise, the best CD found is B-XZZX [Sec.~\ref{sec:generalization}], but the fraction of runs that end on it drops with $p_\mathrm{search}$: 16 of 25 runs at $p_\mathrm{search}=0.1\%$, 6 at $0.25\%$ and none at $0.5\%$.
    Two failure modes contribute.
    First, some runs never sample B-XZZX (7, 13 and 23 runs, respectively [Table~\ref{tab:method_results} in Appendix~\ref{app:results_ci}]) and end on a CD with a higher $P_L$ than B-XZZX, also when evaluated at the elevated $p_\mathrm{search}$, which likely indicates that they are stuck in a local minimum.
    Second, some runs do sample B-XZZX, but end pruning it out at a later stage (2, 6 and 2 runs, respectively [Table~\ref{tab:method_results} in Appendix~\ref{app:results_ci}]).
    Potential fixes to these two problems are discussed in Sec.~\ref{sec:discussion}.

The plots in Fig.~\ref{fig:Results} clearly show that using $p_\mathrm{search} > p$ can reduce the shot count by $1$ to $2$ orders of magnitude.
For the distance-3 surface code under HBD noise, increasing $p_\mathrm{search}$ from $0.1\%=p$ to $0.5\%=5p$ allows the median EBBI run to find the optimal CD $10\times$ faster while random search finds it $4\times$ faster [Fig.~\ref{fig:Results}a].
Meanwhile, under HBD-SI1000 noise, the gain from increasing $p_\mathrm{search}$ is slightly larger for EBBI at about $12\times$, while the median random search run fails to identify the optimal CD under both $p_\mathrm{search}=0.25\%$ and $0.5\%$ [Fig.~\ref{fig:Results}c].
For the distance-5 surface code, the gain increases to close to $80\times$ for EBBI under HBD noise when increasing $p_\mathrm{search}$ from $0.1\%$ to $0.5\%$ [Fig.~\ref{fig:Results}b], with random search never converging due to the exponentially large $3^{25}$ space.
Under HBD-SI1000 noise, increasing $p_\mathrm{search}$ is less successful; at $p_\mathrm{search}=0.25\%$, EBBI almost converges to the B-XZZX code while for $p_\mathrm{search}=0.5\%$ it does not converge at all to the B-XZZX code.
This could either be a result of the lower Pearson correlation under HBD-SI1000 compared to HBD, especially in the top 100 CDs, which was seen in the exhaustive search results for distance-3 [Fig.~\ref{fig:Corr_HBD_MWPM} and Fig.~\ref{fig:Corr_MHBD_MWPM} in Appendix~\ref{app:p_search_corr}] or something caused by the pruning logic in EBBI.
To test this, we used the B-XZZX CD found at $p_\mathrm{search}=p$ and compared its $P_L$ against that of the final CDs found at $p_\mathrm{search}\in\{0.25\%,0.5\%\}$ at their corresponding noise models and saw that B-XZZX remains the CD with the lowest $P_L$.
This leads us to conclude that the failure to converge to the B-XZZX code at $p_\mathrm{search}\in\{0.25\%,0.5\%\}$ is a result of EBBI and not the noise-boosting itself.
These results demonstrate the practical feasibility of increasing $p_\mathrm{search}$ as a method for reducing the required number of shots in circuit-level simulation-based optimization of CDs.

\subsection{Generalization to larger codes}\label{sec:generalization}

A strong dependence between the CD found and the noise model is observed.
In both the distance-3 and distance-5 instances under HBD noise, the well-known XZZX CD is rediscovered by EBBI.
However, when the noise model is changed to HBD-SI1000, XZZX is no longer optimal and a new code, B-XZZX, is discovered, despite the HBD-SI1000 noise model using the same bias $\eta=500$ as that of the HBD noise model and differing only in the relative noise strengths between operations.

\begin{figure}[tb]
    \centering
    \includegraphics[]{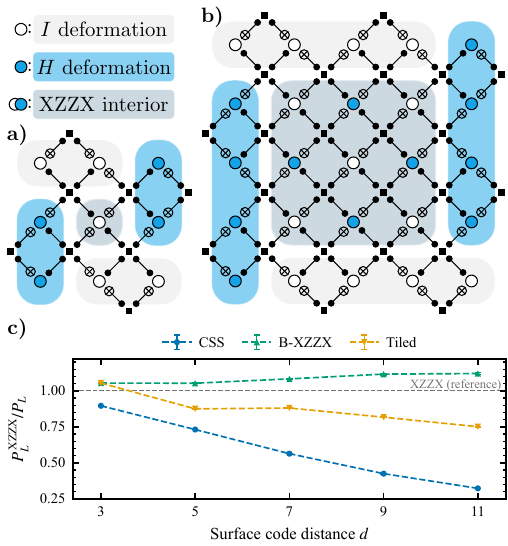}
    \caption{
        Generalization of results.
        \textbf{a)} Optimal distance-3 CD on the surface code under the HBD-SI1000 noise model, obtained from a search over all $3^9$ $\mathcal{A}_3$-CDs; it is also the best CD found by EBBI.
        \textbf{b)} Best distance-5 CD on the surface code under HBD-SI1000 noise found by EBBI.
        \textbf{c)} Improvement in $P_L$ over the XZZX surface code for the undeformed CSS code under HBD-SI1000($p=0.1\%,\eta=500$), the tiled optimal distance-3 CD~\cite{dua2024clifford} (panel a) and the discovered B-XZZX code.
        The error bars show one standard deviation and are too small to see in the plot; each $P_L$ at distance-3 to distance-9 is estimated from $10^{10}$ shots in both the logical X and Z bases of the surface code, with distance-11 using $10^{11}$ shots.
    }
    \label{fig:generalization}
\end{figure}

The best CD found under the HBD-SI1000 noise model for the distance-3 and 5 cases [Fig.~\ref{fig:Results}c,d] can be generalized to larger surface code distances.
It can be viewed as the XZZX CD~\cite{BonillaAtaides2021TheCode} with a modified border.
All the qubits on the left and right edges and on the NE and SW corners use the $H$ deformation, while the remaining edge qubits use no deformation [Fig.~\ref{fig:generalization}a,b].
This generalized form differs from the XZZX code on only $2(d-1)$ qubits in the distance-$d$ case.
For this reason we will refer to it as the bordered XZZX code or simply B-XZZX.

Remarkably, despite the B-XZZX code differing from the XZZX code on only a diminishing fraction $2/d-2/d^2$ of the qubits, its relative performance over the XZZX code increases with code distance [Fig.~\ref{fig:generalization}c].
We note that the improvement over XZZX flattens out between $d=9$ and $d=11$ and could potentially decrease after this point; however, the same thing was observed between $d=3$ and $d=5$ making it hard to conclude anything definitively.
At $d=11$ the $P_L$ under HBD-SI1000 noise with $p=0.1\%$ and $\eta=500$ is $\sim10^{-7}$ making it increasingly expensive to test at even larger code distances.
This shows that despite EBBI only being tested on $d=3$ and $d=5$ surface codes, it is possible to generalize the CD pattern into a scalable CD that maintains good performance.

We speculate that the performance gains of the B-XZZX code under the HBD-SI1000 noise model are a result of the primarily X-errors left on the data qubits by the CX gates [Appendix~\ref{app:CX_bias}].
Interestingly, the B-XZZX code prefers having two CX and one CZ gate on the edge data qubits instead of two CZ and one CX; this choice results in a more even distribution of X and Z errors on the edge data qubits---the gate choice mainly contribute X-errors, while idling of the data qubits during check-qubit measurement mainly contribute Z-errors.

A different proposal for scaling up found CDs on the surface code to larger distances is given in Ref.~\cite{dua2024clifford}.
They propose taking the optimal distance-3 CD from an exhaustive search over all $3^9$ $\mathcal{A}_3$-CDs and tiling the result to obtain larger-distance deformations.
The optimal CD from an exhaustive search yields exactly the distance-3 B-XZZX code; however, tiling it to larger distances under the PyMatching decoder~\cite{Higgott2025sparseblossom} does not preserve its performance advantages [Fig.~\ref{fig:generalization}c].
Already at distance-5, its performance becomes worse than the regular XZZX code.
These results contrast with the findings in Fig.~6 of Ref.~\cite{dua2024clifford} where tiling the code to larger distances continued to give the lowest $P_L$.
We note that our setup differs significantly from that in Ref.~\cite{dua2024clifford} where they used a code-capacity noise model with near-perfect decoding, while here we use a circuit-level noise model with a sub-optimal decoder.
This illustrates the non-trivial problem of generalizing small CDs to larger distances.

\section{Discussion and outlook}\label{sec:discussion}

In this work we have ranked the CDs by directly estimating their $P_L$ through circuit-level simulations, which becomes expensive as the size of the codes increases.
An alternative to this is to construct a cheap surrogate for the performance of a CD and optimize for that instead of $P_L$ directly.
Ref.~\cite{yun2026computationally} has done exactly this, using a decoder-independent surrogate that allows them to find optimal CDs extremely fast, keeping up with an ever-changing calibration map.
However, under both the HBD and HBD-SI1000 circuit-level noise models considered here, the optimal distance-3 surface code $\mathcal{A}_3$-CD changes depending on whether the PyMatching or BP-OSD decoder is used; a Pearson correlation of $\rho=0.88$ and $\rho=0.30$ between the two decoders' $P_L$ over all $\mathcal{A}_3$-CDs is observed for the HBD and HBD-SI1000 noise models respectively [Fig.~\ref{fig:PyMatching_vs_BPOSD} in Appendix~\ref{app:decoder_dependency}].
Thus, to reliably and quickly find the optimal CD for sub-optimal real-time decoders, the surrogate will have to incorporate information about the decoder in addition to information about the noise model and QEC code.

Boundary modifications similar to the ones we perform on the XZZX code to get the B-XZZX code have been applied to the XY code.
Ref.~\cite{PhysRevX.13.031007} noted that while the XY surface code requires $n=d^2$ $Z$-errors for an undetectable logical error, if just a single non-$Z$-error is allowed, an undetectable logical error can be constructed using only $\sqrt{n}-1$ $Z$-errors on the boundary of the code.
They show that if a deformation is performed on every other edge qubit of the code, $\mathcal{O}(\sqrt{n})$ non-$Z$-errors are required to achieve the same effect, significantly increasing the performance of the code [Appendix G of Ref.~\cite{PhysRevX.13.031007}].
However, we believe that what we are seeing in the B-XZZX code is something different as both a logical-$X$ and $Z$ error can be constructed using a single non-$Z$-error and $\sqrt{n}-1$ $Z$-errors.
Thus, the increasing performance of the B-XZZX code over XZZX as the code distance $d$ increases remains unexplained.

In Sec.~\ref{sec:performance} we observed two failure modes of EBBI for the distance-5 surface code under HBD-SI1000 noise.
First, some runs get stuck in a local minimum.
One solution to this is to replace a pruned CD with a uniformly random CD instead of a mutation once many consecutive mutations have been pruned.
This allows EBBI to escape local minima, but may add significant overhead, since a random CD is unlikely to be close to optimal.
Second, some runs discover the optimal CD only to prune it out at a later stage, a result of overly aggressive pruning.
This happens when a new CD, whose $P_L$ estimate is still based on few logical errors, appears better than the optimal CD by chance, causing the optimal CD to satisfy Eq.~\eqref{eq:prune_condition}.
We see three possible solutions.
(i) Decreasing $\gamma$ reduces the rate of such false pruning, at the cost of more shots per CD.
(ii) Increasing $W$ gives a more accurate initial $P_L$ estimate of each new CD, reducing the risk that it falsely appears better than the optimal CD.
(iii) A previously pruned CD could be re-added, in place of a mutation, once it no longer satisfies Eq.~\eqref{eq:prune_condition}.

For device-level application of EBBI, a per-qubit calibration map of the $Z$-bias and of the relative noise strengths between operations and qubits is needed.
While the two noise models considered in this work, HBD and HBD-SI1000, are both uniform noise models, the power of EBBI comes from its use of the strong $P_L$ correlations between similar CDs.
We do not expect that going from a noise model where all qubits are identical to a per-qubit noise model will have any significant effect on this correlation and thus expect that EBBI will continue to perform well under per-qubit noise models calibrated from real devices.
    
While we already see that EBBI performs well, there are several things that might improve its performance even further.
\emph{Pruning:} The way the pruning step currently works is that once a CD is pruned it cannot be added back in at a later point; we have tried removing this restriction, but it resulted in significantly slower convergence.
It would be interesting to explore the possibility of allowing a CD that has previously been pruned out to be re-added under certain conditions.
\emph{Mutation:} The mutation step currently picks random deformations uniformly from $\mathcal{A}$ on a subset of the qubits.
This could potentially be made more efficient by using the cross-entropy method~\cite{rubinstein1999cross,de2005tutorial} for this sampling, so that elements of $\mathcal{A}$ that have worked well in the past are more likely to be used in the future.
In the case of the PyMatching decoder this would decrease the usage of the $H\sqrt{Z}H$ deformation that introduces correlated errors, which we expect would decrease the effective search space from $3^n\rightarrow2^n$ depending on how quickly the $H\sqrt{Z}H$ deformation is ruled out.

\section{Conclusion}

In this work, we have introduced EBBI, a scalable optimization framework for identifying high-performing Clifford-deformed QEC codes under general physical noise models and decoders.
We evaluated the framework using realistic circuit-level noise models suitable for a variety of qubit types and practical decoding based on the minimum weight perfect matching algorithm.
With the surface code as an example case, EBBI identifies CDs overlooked so far where code boundaries are relevant even at large code distances.
Looking ahead, our results establish a CD optimization as a practical tool for tailoring quantum error-correcting codes to hardware-specific noise characteristics that is broadly applicable to various stabilizer codes, decoders, and quantum computer architectures.

\section*{Acknowledgment}

We would like to thank Birk Nøhr Dissing, Jakob Mannstadt, and Eoghan Ryan for helpful discussions during the development of EBBI and interpretation of the results.

This work was supported by the Novo Nordisk Foundation,
Grant number NNF22SA0081175, NNF Quantum Computing Programme.

\section*{Data availability}

The data and code that support the findings of this paper are openly available \cite{EBBI_code}.

\bibliography{references}

\appendix

\section{EBBI Pseudocode}

\begin{algorithm}[H]\label{alg:ebbi}
    \caption{EBBI}
    \SetKwInOut{GInput}{General Inputs}
    \SetKwInOut{MInput}{Method Inputs}
    \GInput{QEC code $\mathcal{C}$; QEC Decoder $\mathcal{D}$; noise model $\mathcal{N}$; shot budget $B$; Single-qubit deformation set $\mathcal{A}$}
    \MInput{number of candidates $K$; Warm-up errors $W$; Prune condition $\gamma$; Mutation probability $\mathcal{M}$}
    \newcommand{\CD}{\mathrm{CD}}
    
    $S_\mathrm{tot} \gets 0$\;
    \For{$k \gets 1$ \KwTo $K$}{
        \tcc{Fig.~\ref{fig:ebbi}b}
        $\CD_k \gets\text{RandomCliffordDeformation}(\mathcal{C}, \mathcal{A})$\;
        $\textit{succ}, \textit{fails} \gets \text{Warmup}(\CD_k, \mathcal{D}, \mathcal{N}, W)$\;
        $\alpha_k \gets \textit{fails} + 1, \quad \beta_k \gets \textit{succ} + 1$\;
        $S_k \gets \textit{succ} + \textit{fails}$\;
        $S_\mathrm{tot} \gets S_\mathrm{tot} + S_k$\;
    }
    
    \While{$S_\mathrm{tot} \le B$}{
        \tcc{Fig.~\ref{fig:ebbi}c}
        $\mathbf{\sigma} \gets \sqrt{\frac{\boldsymbol{\alpha}\boldsymbol{\beta}}{(\boldsymbol{\alpha}+\boldsymbol{\beta})^2(\boldsymbol{\alpha}+\boldsymbol{\beta}+1)}}$\;
        $i \gets \text{argmax}(\mathbf{\sigma})$\;
        $\textit{succ}, \textit{fails} \gets \text{Simulate\&Decode}(\CD_i, \mathcal{D}, \mathcal{N}, S_i)$\;
        $\alpha_i \gets \alpha_i + \textit{fails}, \quad \beta_i \gets \beta_i + \textit{succ}$\;
        $S_i \gets S_i + \textit{succ} + \textit{fails}$\;
        $S_\mathrm{tot} \gets S_\mathrm{tot} + S_i$\;
        
        \tcc{Fig.~\ref{fig:ebbi}d}
        \For{$k \gets 1$ \KwTo $K$}{
            \If{Eq.~\eqref{eq:prune_condition}}{
                $m \gets \operatorname*{argmin}_{m\neq k}\left(\text{BetaRVS}(\boldsymbol{\alpha}, \boldsymbol{\beta})\right)$\;
                $\CD_k \gets \text{Mutate}(\CD_m, \mathcal{A}, \mathcal{M})$\;
                $\textit{succ}, \textit{fails} \gets \text{Warmup}(\CD_k, \mathcal{D}, \mathcal{N}, W)$\;
                $\alpha_k \gets \textit{fails} + 1, \quad \beta_k \gets \textit{succ} + 1$\;
                $S_k \gets \textit{succ} + \textit{fails}$\;
                $S_\mathrm{tot} \gets S_\mathrm{tot} + S_k$\;
            }
        }
    }
    \tcc{Return the currently best known CD}
    $\boldsymbol{\mu} \gets \frac{\boldsymbol{\alpha}}{\boldsymbol{\alpha} + \boldsymbol{\beta}},\quad\boldsymbol{\sigma} \gets \sqrt{\frac{\boldsymbol{\alpha}\boldsymbol{\beta}}{(\boldsymbol{\alpha}+\boldsymbol{\beta})^2(\boldsymbol{\alpha}+\boldsymbol{\beta}+1)}}$\;
    $k^* \gets \operatorname*{argmin}(\boldsymbol{\mu} + \boldsymbol{\sigma})$\;
    \Return $\CD_{k^*}$\;
\end{algorithm}

\section{Noise model expansions}\label{app:noise_details}

Description of the noise channels $\mathcal{E}$ used in Table~\ref{tab:noise_models} as a quantum channels following the definition from Ref.~\cite{etxezarreta2026leveraging}.
In the following $\rho=|\psi\rangle\langle\psi|$ is the density matrix of the quantum state of the system $|\psi\rangle$ and $X$, $Y$ and $Z$ are the Pauli gates with $I$ the identity.

\begin{align}
\mathcal{E}^{\mathrm{DEP}}_{p}(\rho) &= (1-p)\rho + \frac{p}{3}\left(X\rho X + Y\rho Y + Z\rho Z\right) \\\nonumber\\
\mathcal{E}^{X}_{p}(\rho) &= (1-p)\rho + p\,X\rho X                                                    \\\nonumber\\
\mathcal{E}^{\mathrm{Bias}_1}_{p,\eta}(\rho) &= (1-p)\rho                                                 \nonumber
\\&+ \frac{p}{2(\eta+1)}\left(X\rho X + Y\rho Y\right)                                                    \nonumber
\\&+ \frac{\eta p}{\eta+1}\,Z\rho Z                                                                     \\\nonumber\\
\mathcal{E}^{\mathrm{Bias}_2}_{p,\eta}(\rho) &= (1-p)\rho                                               \nonumber
\\&+ \frac{\eta p}{3(\eta+1)}\!\!\sum_{P\in\{IZ,ZI,ZZ\}}\!\! P\rho P^{\dagger}                            \nonumber
\\&+ \frac{p}{12(\eta+1)}\!\!\sum_{P\in\mathcal{P}_2\setminus\{II,IZ,ZI,ZZ\}}\!\! P\rho P^{\dagger}
\end{align}

\noindent with $\mathcal{P}_2=\{I,X,Y,Z\}^{\otimes2}$ the set of the 16 two-qubit Pauli strings.

\paragraph*{Noiseless operations} While most operations (initialization, gates, idling, and measurements) are noisy, there are a few places where we implement them perfectly.
During the basis change from $|00\cdots0\rangle$ to the Clifford-deformed logical X or Z projection, we use noise-less single qubit gates.
The same goes for the basis change right before logical readout.
We do this to ensure that the noise model between the X and Z memory experiment places the Pauli-errors at the exact same locations in the circuit under the same seed, independent of whether the logical X and Z basis is used.
This is what enables us to infer $P_L^X$, $P_L^Y$, $P_L^Z$, instead of $P_L^X+P_L^Y$ and $P_L^Z+P_L^Y$ for the X and Z memory experiments respectively, that we would otherwise get when the error locations are not the same.

\section{Circuit construction}\label{app:circuit_construction}

\begin{figure*}[tb]
    \includegraphics[width=\linewidth]{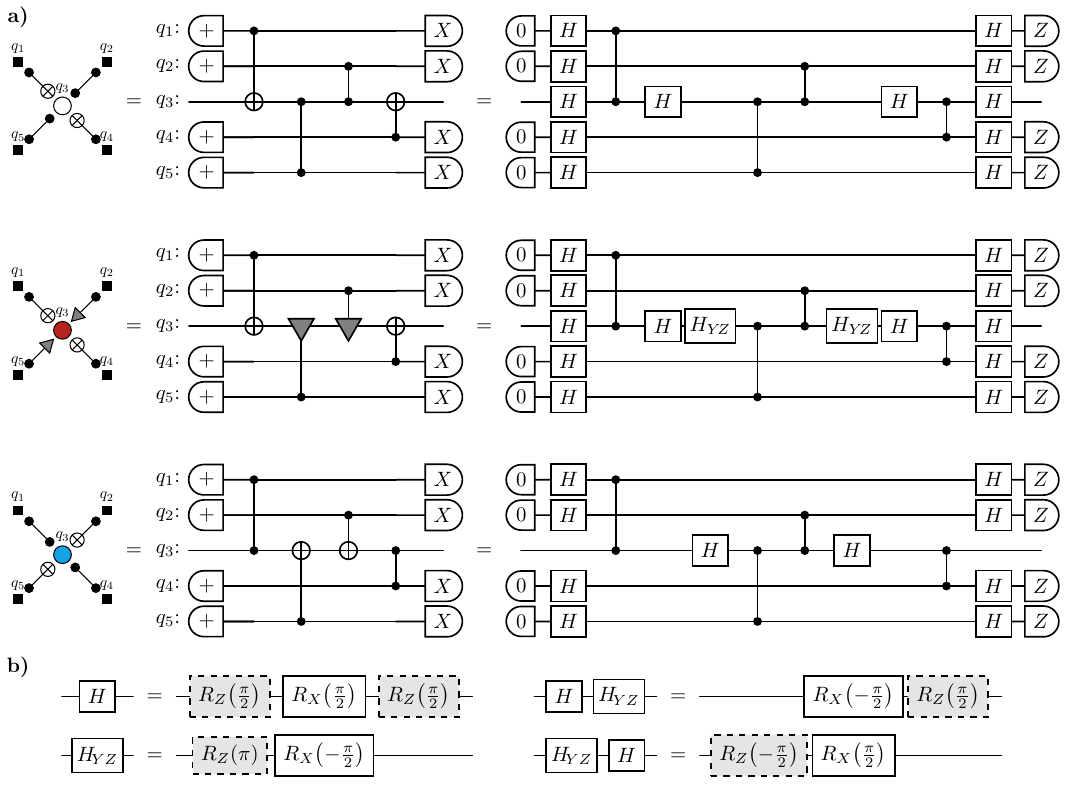}
    \caption{
        \textbf{a)} Circuit diagrams illustrating stabilizer measurements in the surface code under different deformations in $\mathcal{A}_3$.
        \emph{Left:} Measurement circuits expressed using the corresponding controlled-Pauli interactions.
        \emph{Right:} Controlled-$X$ and controlled-$Y$ operations are decomposed into a single controlled-$Z$ gate together with basis changes on the data qubit.
        \textbf{b)} $ZXZ$ Euler decomposition up to a global phase of the basis-change gates used in panel a and their products.
        Virtual $R_Z$ rotations (dashed boxes) are implemented in software and therefore introduce negligible latency and are thus treated as ideal in this work.
    }
    \label{fig:circuit_decomposition} 
\end{figure*}

Circuit construction can be viewed in three levels.
(1) The circuit is constructed using all three two-qubit controlled gates (CX, CY and CZ) and check qubits are initialized and measured in the X-basis.
This is the most direct construction from the code schematics [Fig.~\ref{fig:ebbi}a, Fig.~\ref{fig:circuit_decomposition}a left].
(2) Measurement and initialization can be compiled to the Z-basis using $H$ gates and the CX and CY gates can be constructed with $CX = (I\otimes H)CZ(I\otimes H)$ and $CY = (I\otimes H_{YZ})CZ(I\otimes H_{YZ})$ [Fig.~\ref{fig:circuit_decomposition}a right].
(3) Finally, the single-qubit gates $H$ and $H_{YZ}$, along with their products $HH_{YZ}$ and $H_{YZ}H$ can be Euler decomposed into a $Z-X-Z$ rotation [Fig.~\ref{fig:circuit_decomposition}b].
The two $Z$ rotations are usually handled in the control software without requiring any signals to be sent to the quantum chip, making them instantaneous and error free.
For this reason, the $H$, $H_{YZ}$, $HH_{YZ}$, and $H_{YZ}H$ can all be viewed as a single $\pi/2$ X-rotation and be given the exact same single-qubit depolarization error channel.

\subsection{Bias of CX and CY gates}\label{app:CX_bias}

This choice of compilation of CX and CY gates using single-qubit rotations and CZ gates leads to a bias on the CX and CY gates.
Going from our noise model [Table~\ref{tab:noise_models}, Appendix~\ref{app:noise_details}], the CZ gates primarily produce Z-errors on both target and control qubits under high bias $\eta\gg1$.
The control qubit (check qubit in the case of QEC) remains unmodified, maintaining a dominant Z-error term, while the target qubits (data qubits) transform the Z-error into $X$-errors ($Y$-errors) after the application of $H$ ($H_{YZ}$) gates for CX (CY) gate.
If the single-qubit rotations have low error rates, which is the case under the HBD-SI1000 noise model, the target qubit (data qubit) will be heavily biased in the $X$ ($Y$) direction for the CX (CY) gate.
However, under the HBD noise model where the noise strength of the single-qubit operations is the same as that of the two-qubit CZ gate, this effect is washed out due to the large depolarization channel.

\section{Reducing the Clifford-deformation set in the circuit-level noise model regime}\label{app:cd_reduction}

Here we give an informal argument for why it is sufficient to test $\mathcal{A}_3$ deformation set and not the full $\mathcal{A}_6$ deformation set under both the HBD and HBD-SI1000 noise models; confirmed empirically by comparing detector error models for the respective circuits.

The full CD set can be written in terms of the reduced set as $\mathcal{A}_6=\mathcal{A}_3\cup\sqrt{Z}\mathcal{A}_3$ where equality holds up to Pauli factors that only flip stabilizer signs and therefore do not affect $P_L$.
In the code-capacity and phenomenological noise models, this is immediate: $\sqrt{Z}$ swaps the $p_X$ and $p_Y$ channels, so assuming $p_X=p_Y$ results in identical noise and thus identical $P_L$.

At the circuit level using HBD or HBD-SI1000 noise, a similar thing is going on.
As for the data idling errors, the argument used in the code-capacity noise model carries over directly.
The $X$-errors on initialization and measurements remain unaffected by the CD as they commute with the $CZ$ gates (after being turned into $Z$-errors by the $H$ gates on the check qubits) and result in a classical readout error, just like in the phenomenological noise model case.

As for the errors that occur during the stabilizer measurement:
The $\sqrt{Z}$ swaps $CX$ and $CY$ two-qubit gates, which after circuit compilation [Appendix~\ref{app:circuit_construction}] leads to swapping the $H$ and $H_{YZ}$ single-qubit gates; leaving the gate count and ordering unchanged.
Both the HBD and HBD-SI1000 noise models apply $\mathrm{DEP}$ errors on both of these gates, which from its $p_X=p_Y=p_Z$ symmetry leaves the detector error model unchanged.
The only non-uniform noise channel occurs after the $CZ$ gate, which primarily produces $Z$ errors that will change into $X$ or $Y$ after the $H$ or $H_{YZ}$ gates respectively, depending on whether a $\sqrt{Z}$ deformation was applied.
However, this has no net effect on the detector error model as the commutation relations with the other detectors in the code are unchanged (any stabilizer acting on the given data qubit is transformed the same way as the error, so whether they commute is unchanged).
As a result, applying a $\sqrt{Z}$ deformation on any data qubit in the code will have no net effect on the detector error model and thus not change $P_L$, independent of which decoder is used.

To confirm that this holds in practice, Stim detector error models have been compared.
Two circuits are created; the first circuit uses a random uniformly drawn CD from $\mathcal{A}_6^n$ while the second circuit modifies the CD by applying $\sqrt{Z}$ to a subset of the qubits at random.
The detector error models from the two circuits are then compared using Stim's \texttt{dem1.approx\_equals(dem2, atol=1e-15)}.
This was done on both the distance-3 and distance-5 surface codes, 1000 times each, under either HBD or HBD-SI1000 noise and in both the logical X and Z-basis.
An exact match in the detector error models was observed for all cases used in this paper, confirming empirically the validity of using EBBI on the $\mathcal{A}_3$ set instead of the full $\mathcal{A}_6$ set.

\section{Exhaustive search}\label{app:exhaustive}

\begin{figure}[tb]
    \centering
    \includegraphics[]{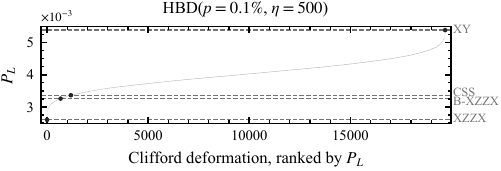}
    \caption{
        Rank-plot of the $P_L$ of the $3^9$ distance-3 surface code $\mathcal{A}_3$-CDs under HBD($p=0.1\%$, $\eta=500$) noise.
    }
    \label{fig:Exhuastive_HBD_MWPM}
\end{figure}

\begin{figure}[tb]
    \centering
    \includegraphics[]{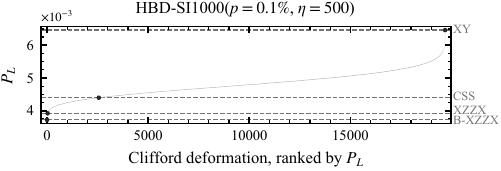}
    \caption{
        Rank-plot of the $P_L$ of the $3^9$ distance-3 surface code $\mathcal{A}_3$-CDs under HBD-SI1000($p=0.1\%$, $\eta=500$) noise.
    }
    \label{fig:Exhuastive_MHBD_MWPM}
\end{figure}

As the number of distance-3 surface code $\mathcal{A}_3$-CDs is small, $3^9=19\,683$, an exhaustive search is a feasible option and can be used to verify the findings from EBBI.
Fig.~\ref{fig:Exhuastive_HBD_MWPM} and Fig.~\ref{fig:Exhuastive_MHBD_MWPM} show the results of an exhaustive search using PyMatching over the HBD and HBD-SI1000 noise models, respectively, with $p=0.1\%$ and $\eta=500$, matching the conditions that EBBI and random search were benchmarked against in Fig.~\ref{fig:Results}.
The CDs run along the X-axis and have been sorted by their $P_L$; the four named CDs: CSS, XZZX, XY and B-XZZX, have also been highlighted.
These plots clearly show that a sharp improvement in the CD's $P_L$ only happens in the top fraction of the space, matching the code-capacity results from Fig.~2a in Ref.~\cite{dua2024clifford}.

\subsection*{Impact from scaling physical error rates}\label{app:p_search_corr}

\begin{figure}[tb]
    \centering
    \includegraphics[]{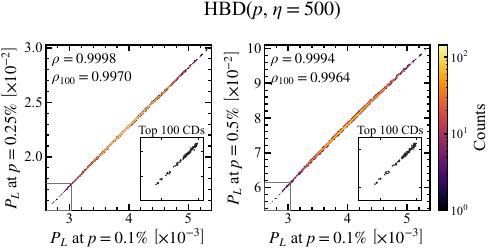}
    \caption{
        Pearson correlation of the CD $P_L$ as the HBD noise model error strength $p$ is increased from $p=0.1\%$ to $p=0.25\%$ (left plot) and $p=0.5\%$ (right plot).
        Correlation numbers are reported for both the entire CD space and for only the 100 CDs with the lowest $P_L$ (at $p=0.1\%$) shown in the insets.
        Each $P_L$ was estimated using $10^8$ shots.
    }
    \label{fig:Corr_HBD_MWPM}
\end{figure}

\begin{figure}[tb]
    \centering
    \includegraphics[]{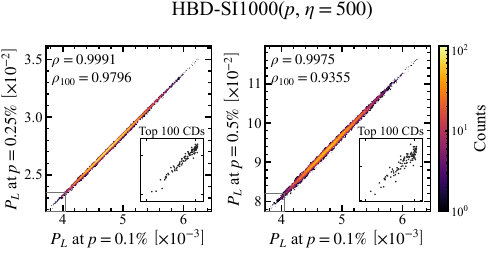}
    \caption{
        Same as Fig.~\ref{fig:Corr_HBD_MWPM} but using the HBD-SI1000 noise model instead.
    }
    \label{fig:Corr_MHBD_MWPM}
\end{figure}

In the main text we make the claim that scaling the physical error rate $p$ only has a small effect on the CD ordering and is thus a practical way of reducing shot counts for both EBBI and random search.
Here we do an exhaustive search using PyMatching over both the HBD and HBD-SI1000 noise models, with $\eta=500$ and $p\in\{0.1\%,0.25\%,0.5\%\}$, matching the settings used in Fig.~\ref{fig:Results}.
Fig.~\ref{fig:Corr_HBD_MWPM} and Fig.~\ref{fig:Corr_MHBD_MWPM} show the correlation under HBD and HBD-SI1000 noise respectively, between the base $p=0.1\%$ and the elevated $p=0.25\%$ and $p=0.5\%$, over both all $3^9$ $\mathcal{A}_3$-CD set, and over the top 100 CDs (the 100 CDs with the lowest $P_L$ under $p=0.1\%$).
We see that the correlation holds very strongly for the HBD noise model, and decreases slightly for the HBD-SI1000 noise model.

\subsection*{Correlation}\label{app:corr}

\begin{figure}[tb]
    \centering
    \includegraphics[]{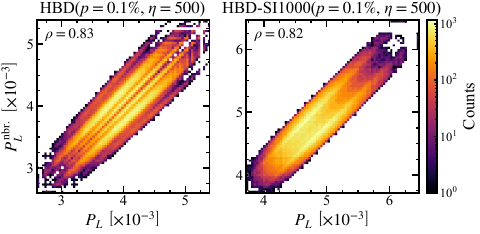}
    \caption{
        Correlation plots between $P_L$ and $P_L^\text{nbr.}$ between two CDs that differ on only a single qubit on the distance-3 surface code.
        \textbf{Left:} using the HBD noise model, \textbf{Right:} using the HBD-SI1000 noise model, both with $p=0.1\%$ and $\eta=500$.
    }
    \label{fig:corr}
\end{figure}

Lastly, we use the exhaustive search results for PyMatching under $p=0.1\%$ and $\eta=500$ for both noise models to check the correlation in the CD's $P_L$ between two CDs that only differ in deformation on only a single qubit.
We call a CD that differs from a different CD on only a single qubit a neighbor and denote its $P_L$ as $P_L^\mathrm{nbr.}$.
In the distance-3 surface code $\mathcal{A}_3$-CD case with 3 possible deformations per qubit, each qubit can be changed to two other deformations and we have 9 data qubits, thus giving each CD 18 neighbors.
The results are shown in Fig.~\ref{fig:corr}.

\section{Random Search}\label{app:random search}

Random search chooses uniformly random CDs of the QEC code and samples each a fixed number of times to estimate their $P_L$.
Random search keeps track of only the CD with the lowest estimated $P_L$.
It is initialized with a fixed shot budget $B$ along with how many shots $S_\mathrm{CD}$ to spend per CD, meaning it will sample $B/S_\mathrm{CD}$ random CDs.

This method does not scale well with codes of increasing size due to two factors: (1) the number of possible CDs grows exponentially as $|\mathcal{A}|^n$ in the number of data qubits $n$ so the number of tested CDs will have to also grow exponentially to keep up; (2) as the code distance $d$ increases, $P_L$ decreases, increasing the number of required shots per CD as $\mathcal{O}\left(1/P_L\right)$.
It could be argued that we only care about the top $q$ percentile and thus only need $\mathcal{O}\left(q^{-1}\right)$ samples; however, for the distance-3 surface code under circuit-level noise we see that there is a sharp drop in $P_L$ close to the $q$ percentile [Fig.~\ref{fig:Exhuastive_HBD_MWPM},\ref{fig:Exhuastive_MHBD_MWPM}].
This is similar to what was seen in Fig.~2a in Ref.~\cite{dua2024clifford} for the code-capacity case.
The second point could be mitigated by artificially increasing the physical qubit error rate to near threshold, which was observed to help but not solve the problem entirely [Fig.~\ref{fig:Results}].

\begin{algorithm}[H]
    \caption{Random Search}
    \SetKwInOut{GInput}{General Inputs}
    \SetKwInOut{MInput}{Method Inputs}
    \SetKw{KwBreak}{break}
    \GInput{QEC code $\mathcal{C}$; QEC Decoder $\mathcal{D}$; noise model $\mathcal{N}$; shot budget $B$}
    \MInput{Shots per CD $S_\mathrm{CD}$}
    \newcommand{\CD}{\mathrm{CD}}

    $P_L^* \gets 1$\tcc*{best seen $P_L$}
    $\mathrm{used\_cd} \gets \emptyset$\;
    \For{$i \gets 1$ \KwTo $\lfloor B/S_\mathrm{CD} \rfloor$}{
        \For{$j \gets 1$ \KwTo $100$}{
            $\CD \gets \text{RandomCliffordDeformation}(\mathcal{C})$\;
            \If{$\CD \notin \mathrm{used\_cd}$}{
                $\mathrm{used\_cd} \gets \mathrm{used\_cd} \cup \{\CD\}$\;
                \KwBreak\;
            }
            \If{$j = 100$}{
                \Return $\CD_\mathrm{best}$\;
            }
        }
        $P_L \gets \text{EstimateLER}(\CD, \mathcal{D}, \mathcal{N}, S_\mathrm{CD})$\;
        \If{$P_L < P_L^*$}{
            $P_L^* \gets P_L$\;
            $\CD_\mathrm{best} \gets \CD$\;
        }
    }
    \Return $\CD_\mathrm{best}$\;
\end{algorithm}

\section{EBBI and random search run spread}\label{app:results_ci}

Fig.~\ref{fig:Results} showed the median $P_L$ over 25 runs for each of the 24 settings: two code distances, two noise models, two methods, and three $p_\mathrm{search}$ values.
In Fig.~\ref{fig:results_ci}, these settings are split into 12 plots instead of the 4 used in Fig.~\ref{fig:Results}, one for each combination of code distance, noise model and $p_\mathrm{search}$.
This allows us to show both the full spread in $P_L$ over all 25 runs and the spread over the central 15 runs (20th to 80th percentile).
Additionally, Table~\ref{tab:method_results} shows, for each setting, how many runs ended at the optimal CD (or what we believe is the optimal CD in the distance-5 case), how many runs found the optimal CD at some point but did not end with it as their final output, and how many runs never found the optimal CD.

The spread confirms that the medians in Fig.~\ref{fig:Results} are representative of the individual runs.
For the distance-3 surface code under both noise models, and for the distance-5 surface code under HBD noise, nearly all EBBI runs end at or near the optimal CD, and the EBBI spread is narrow compared to that of random search.
The exception is the distance-5 surface code under HBD-SI1000 noise, where the number of EBBI runs ending at B-XZZX drops from 16 to 0 as $p_\mathrm{search}$ increases from $0.1\%$ to $0.5\%$, with most of the remaining runs never sampling B-XZZX.

\begin{table}[tb]
    \centering
    \begin{ruledtabular}
    \begin{tabular}{cc|cc|cc}
        \multirow{3}{*}{} & \multirow{3}{*}{$p_\mathrm{search}$} & \multicolumn{2}{c|}{$d=3$ surface code} & \multicolumn{2}{c}{$d=5$ surface code} \\
        \cline{3-6}
         &  & HBD & HBD-SI1000 & HBD & HBD-SI1000 \\
         &  & (XZZX) & (B-XZZX) & (XZZX) & (B-XZZX) \\
        \hline
        \multirow{3}{*}{EBBI} & $0.1\%$ & 25/0/0 & 25/0/0 & 23/2/0 & 16/2/7 \\
         & $0.25\%$ & 25/0/0 & 24/0/1 & 22/2/1 & 6/6/13 \\
         & $0.5\%$ & 25/0/0 & 25/0/0 & 25/0/0 & 0/2/23 \\
        \hline
        \multirow{3}{*}{\begin{tabular}{@{}c@{}}Random\\search\end{tabular}} & $0.1\%$ & 19/0/6 & 15/3/7 & 0/0/25 & 0/0/25 \\
         & $0.25\%$ & 21/0/4 & 9/4/12 & 0/0/25 & 0/0/25 \\
         & $0.5\%$ & 20/0/5 & 9/4/12 & 0/0/25 & 0/0/25 \\
    \end{tabular}
    \end{ruledtabular}
    \caption{
        Result of each of the 25 runs per setting in the form \textbf{a}/\textbf{b}/\textbf{c}.
        \textbf{a}: Runs that ended at the optimal CD (XZZX for HBD and B-XZZX for HBD-SI1000).
        \textbf{b}: Runs that at some point found the optimal CD but then pruned it.
        \textbf{c}: Runs that never found the optimal CD at any point.
    }
    \label{tab:method_results}
\end{table}

\section{Optimal CD depends on decoder}\label{app:decoder_dependency}

\begin{figure}[t]
    \centering
    \includegraphics{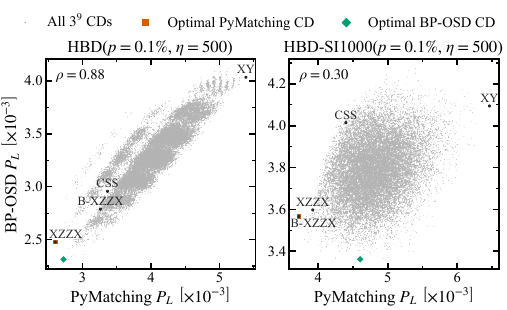}
    \caption{
        Pearson correlation of the $P_L$ of all $3^9$ $\mathcal{A}_3$-CDs between the PyMatching and BP-OSD decoders with the optimal CD for either decoder marked in addition to the named CDs in this paper.
        the $P_L$ were estimated using $10^8$ shots for PyMatching and $10^7$ shots for BP-OSD due to computational cost.
        \textbf{Left:} uses the HBD noise model with $p=0.1\%$ and $\eta=500$.
        \textbf{Right:} uses the HBD-SI1000 noise model with the same parameters as HBD.
    }
    \label{fig:PyMatching_vs_BPOSD}
\end{figure}

\begin{figure*}[tb]
    \centering
    \includegraphics{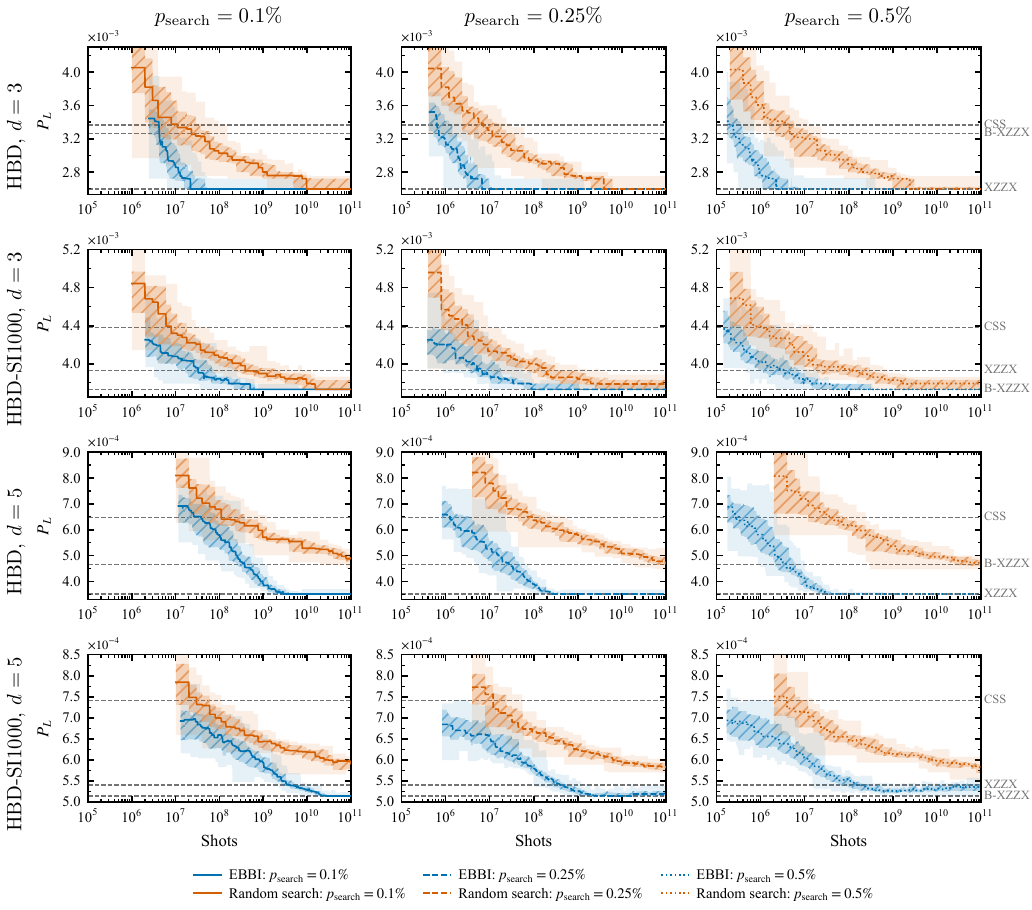}
    \caption{
        Same results as shown in Fig.~\ref{fig:Results}, but with each plot split into a row of 3 plots, allowing for additional per-run details.
        Lightly shaded region shows the full spread over all 25 runs.
        Heavily shaded hashed region shows the spread of the central 15 runs (20th to 80th percentile).
    }
    \label{fig:results_ci}
\end{figure*}

To test to which degree the choice of decoder impacts which CD is optimal, we compare PyMatching decoding against BP-OSD decoding on the $\mathcal{A}_3$-CDs.
An exhaustive search over all $\mathcal{A}_3$-CDs using either PyMatching or BP-OSD for both the HBD and HBD-SI1000 noise models, at $p=0.1\%$ and $\eta=500$, is performed.
In the HBD case, we see that the ordering remains somewhat similar, with a Pearson correlation of $\rho=0.88$, but that the optimal CD has changed to a different one [Fig.~\ref{fig:PyMatching_vs_BPOSD} left].
In the HBD-SI1000 case, the difference in ordering becomes much stronger with the Pearson correlation dropping down to $\rho=0.30$, making the two CD orderings almost independent of each other [Fig.~\ref{fig:PyMatching_vs_BPOSD} right].
This highlights the importance of the role the decoder plays in determining optimal or close to optimal CDs.

\begin{figure*}[tb]
    \centering
    \includegraphics{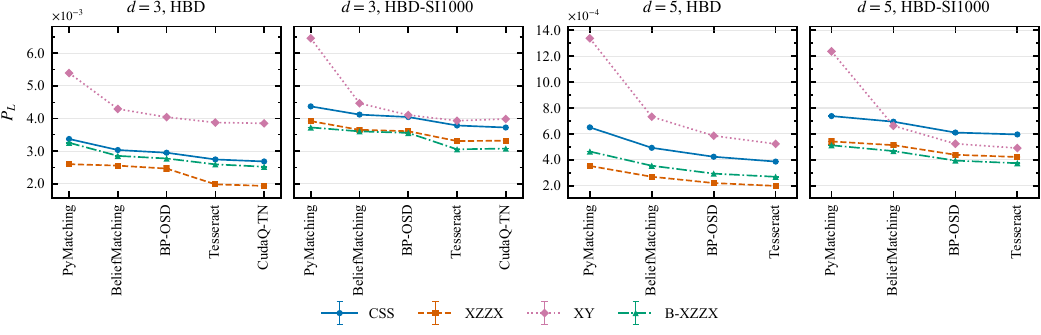}
    \caption{
        The effect on $P_L$ (estimated from $10^8$ shots) induced by the decoder across 4 CDs on the distance-3 and 5 surface code under either HBD or HBD-SI1000 noise.
        Both noise models use $p=0.001$ and $\eta=500$.
        The error bars on the data are too small to see and hidden behind the markers.
        The raw data for each data point can be found in Table~\ref{tab:decoder_dependence}.
        The following decoder abbreviations are used: PyMatching~\cite{Higgott2025sparseblossom}, BeliefMatching~\cite{PhysRevX.13.031007}, BP-OSD~\cite{roffe_decoding_2020}, Tesseract~\cite{beni2025tesseractdecoder} and the CudaQ Tensor Network (CudaQ-TN) decoder~\cite{cudaq, cudaqx}.
    }
    \label{fig:decoder_ranking}
\end{figure*}

In addition to the PyMatching and BP-OSD decoders, we also test the BeliefMatching, Tesseract and CudaQ-TN decoders on the four named CDs in this paper, CSS, XZZX, XY, and B-XZZX.
Fig.~\ref{fig:decoder_ranking} shows the results of these tests, with the exact numbers reported in Table~\ref{tab:decoder_dependence}.
This data shows that especially the XY CD's $P_L$ can be significantly reduced by switching to a better decoder while the other CDs see smaller improvements.
The runtime for each decoder is listed in Table~\ref{tab:decoder_runtime}.
The four CPU decoders, PyMatching, BeliefMatching, BP-OSD and Tesseract, only ran on a single CPU core, which gives plenty of opportunity for parallelization in the future, which would allow EBBI and random search to either use better decoders or run on larger QEC codes.
The CudaQ-TN decoder on the other hand uses a full GPU and will thus be much harder to speed up.

\begin{table*}[tb]
    \centering
    \begin{ruledtabular}
    \begin{tabular}{ccc|cccc}
        Distance & Noise model & Decoder & CSS & XZZX & XY & B-XZZX \\
        \hline
        \multirow{10}{*}{3} & \multirow{5}{*}{HBD} & PyMatching & $3.3731(58) \times 10^{-3}$ & $\mathbf{2.6005(51) \times 10^{-3}}$ & $5.3951(73) \times 10^{-3}$ & $3.2569(57) \times 10^{-3}$ \\
         &  & BeliefMatching & $3.0349(55) \times 10^{-3}$ & $\mathbf{2.5559(50) \times 10^{-3}}$ & $4.2939(65) \times 10^{-3}$ & $2.8547(53) \times 10^{-3}$ \\
         &  & BP-OSD & $2.9517(54) \times 10^{-3}$ & $\mathbf{2.4657(50) \times 10^{-3}}$ & $4.0424(63) \times 10^{-3}$ & $2.7748(53) \times 10^{-3}$ \\
         &  & Tesseract & $2.7454(52) \times 10^{-3}$ & $\mathbf{1.9830(44) \times 10^{-3}}$ & $3.8773(62) \times 10^{-3}$ & $2.5950(51) \times 10^{-3}$ \\
         &  & CudaQ-TN & $2.6836(52) \times 10^{-3}$ & $\mathbf{1.9344(44) \times 10^{-3}}$ & $3.8531(62) \times 10^{-3}$ & $2.5228(50) \times 10^{-3}$ \\
        \cline{2-7}
         & \multirow{5}{*}{HBD-SI1000} & PyMatching & $4.3723(66) \times 10^{-3}$ & $3.9204(62) \times 10^{-3}$ & $6.4594(80) \times 10^{-3}$ & $\mathbf{3.7263(61) \times 10^{-3}}$ \\
         &  & BeliefMatching & $4.1204(64) \times 10^{-3}$ & $3.6532(60) \times 10^{-3}$ & $4.4681(67) \times 10^{-3}$ & $\mathbf{3.6063(60) \times 10^{-3}}$ \\
         &  & BP-OSD & $4.0457(63) \times 10^{-3}$ & $3.6107(60) \times 10^{-3}$ & $4.1056(64) \times 10^{-3}$ & $\mathbf{3.5602(60) \times 10^{-3}}$ \\
         &  & Tesseract & $3.7870(61) \times 10^{-3}$ & $3.3101(57) \times 10^{-3}$ & $3.9329(63) \times 10^{-3}$ & $\mathbf{3.0559(55) \times 10^{-3}}$ \\
         &  & CudaQ-TN & $3.7244(61) \times 10^{-3}$ & $3.3202(58) \times 10^{-3}$ & $3.9858(63) \times 10^{-3}$ & $\mathbf{3.0794(55) \times 10^{-3}}$ \\
        \hline
        \multirow{8}{*}{5} & \multirow{4}{*}{HBD} & PyMatching & $6.508(26) \times 10^{-4}$ & $\mathbf{3.535(19) \times 10^{-4}}$ & $1.3368(37) \times 10^{-3}$ & $4.646(22) \times 10^{-4}$ \\
         &  & BeliefMatching & $4.933(22) \times 10^{-4}$ & $\mathbf{2.705(16) \times 10^{-4}}$ & $7.323(27) \times 10^{-4}$ & $3.552(19) \times 10^{-4}$ \\
         &  & BP-OSD & $4.245(21) \times 10^{-4}$ & $\mathbf{2.219(15) \times 10^{-4}}$ & $5.859(24) \times 10^{-4}$ & $2.945(17) \times 10^{-4}$ \\
         &  & Tesseract & $3.872(20) \times 10^{-4}$ & $\mathbf{1.999(14) \times 10^{-4}}$ & $5.234(23) \times 10^{-4}$ & $2.700(16) \times 10^{-4}$ \\
        \cline{2-7}
         & \multirow{4}{*}{HBD-SI1000} & PyMatching & $7.377(27) \times 10^{-4}$ & $5.422(23) \times 10^{-4}$ & $1.2364(35) \times 10^{-3}$ & $\mathbf{5.126(23) \times 10^{-4}}$ \\
         &  & BeliefMatching & $6.945(26) \times 10^{-4}$ & $5.139(23) \times 10^{-4}$ & $6.628(26) \times 10^{-4}$ & $\mathbf{4.682(22) \times 10^{-4}}$ \\
         &  & BP-OSD & $6.108(25) \times 10^{-4}$ & $4.397(21) \times 10^{-4}$ & $5.244(23) \times 10^{-4}$ & $\mathbf{3.949(20) \times 10^{-4}}$ \\
         &  & Tesseract & $5.963(24) \times 10^{-4}$ & $4.226(21) \times 10^{-4}$ & $4.913(22) \times 10^{-4}$ & $\mathbf{3.747(19) \times 10^{-4}}$ \\
    \end{tabular}
    \end{ruledtabular}
    \caption{
        Estimated $P_L$ from $10^8$ shots with $p=0.1\%$ and $\eta=500$ on the CSS, XZZX, XY and B-XZZX CDs.
        Bold entries indicate the CD with the lowest recorded $P_L$.
    }
    \label{tab:decoder_dependence}
\end{table*}

\begin{table*}[tb]
    \centering
    \begin{ruledtabular}
    \begin{tabular}{l|cc|cc|l}
        \multirow{2}{*}{Decoder} & \multicolumn{2}{c|}{$d = 3$} & \multicolumn{2}{c|}{$d = 5$} & \multirow{2}{*}{Compute resource} \\
         & HBD & HBD-SI1000 & HBD & HBD-SI1000 &  \\
        \hline
        PyMatching & $25.5\,\mathrm{s}$ ($1\times$) & $32.8\,\mathrm{s}$ ($1\times$) & $6\,\mathrm{m}\,34\,\mathrm{s}$ ($1\times$) & $9\,\mathrm{m}\,01\,\mathrm{s}$ ($1\times$) & 1 core, AMD EPYC 9475F \\
        BeliefMatching & $53\,\mathrm{m}\,06\,\mathrm{s}$ ($125\times$) & $1\,\mathrm{h}\,01\,\mathrm{m}$ ($113\times$) & $22\,\mathrm{h}\,20\,\mathrm{m}$ ($204\times$) & $22\,\mathrm{h}\,06\,\mathrm{m}$ ($147\times$) & 1 core, AMD EPYC 9475F \\
        BP-OSD & $1\,\mathrm{h}\,11\,\mathrm{m}$ ($168\times$) & $1\,\mathrm{h}\,23\,\mathrm{m}$ ($153\times$) & $45\,\mathrm{h}\,09\,\mathrm{m}$ ($413\times$) & $42\,\mathrm{h}\,53\,\mathrm{m}$ ($285\times$) & 1 core, AMD EPYC 9475F \\
        Tesseract & $1\,\mathrm{h}\,49\,\mathrm{m}$ ($258\times$) & $2\,\mathrm{h}\,53\,\mathrm{m}$ ($317\times$) & $27\,\mathrm{h}\,03\,\mathrm{m}$ ($247\times$) & $40\,\mathrm{h}\,15\,\mathrm{m}$ ($268\times$) & 1 core, AMD EPYC 9475F \\
        CudaQ-TN & $41\,\mathrm{m}\,12\,\mathrm{s}$ ($96.9\times$) & $41\,\mathrm{m}\,06\,\mathrm{s}$ ($75.3\times$) & --- & --- & 1 NVIDIA H200 NVL, 141 GB \\
    \end{tabular}
    \end{ruledtabular}
    \caption{
        Runtime of the different decoders tested on both HBD and HBD-SI1000 noise with $p=0.1\%$ and $\eta=500$ using $10^8$ shots.
        The numbers in parentheses show how much faster PyMatching is compared to the given decoder, noise model and code distance.
    }
    \label{tab:decoder_runtime}
\end{table*}

\end{document}